High entropy alloys and their affinity to hydrogen: from Cantor to platinum group elements alloys.

K. Glazyrin,[a] K. Spektor,[a] M. Bykov,[b] W. Dong,[a] J.-H. Yu,[c] S. Yang,[c] J.-S. Lee,[d] S. Divinski,[e] M. Hanfland,[f] K. V. Yusenko[g]

[a] *Photon Sciences, Deutsches Elektronen-Synchrotron, Notkestr. 85, 22607 Hamburg, Germany*
[b] *Institute of Inorganic Chemistry, University of Cologne, Cologne, Germany*
[c] *Powder Materials Division, Korea Institute of Materials Science, 51508 Changwon, South Korea*
[d] *Department of Materials Science and Chemical Engineering, Hanyang University, 15588 Ansan, South Korea*
[e] *Institute of Materials Physics, University of Münster, D-48149 Münster, Germany*
[f] *ESRF – The European Synchrotron, 71 Av. des Martyrs, 38000, Grenoble, France*
[g] *Bundesanstalt für Materialforschung und –prüfung (BAM), D-12489 Berlin, Germany*

**Abstract**
Properties of high entropy alloys are currently in the spotlight due to their promising applications. One of the least investigated aspects is the affinity of these alloys to hydrogen, its diffusion and reactions. In this study we apply high-pressure at ambient temperature and investigate stress-induced diffusion of hydrogen into the structure of high entropy alloys (HEA) including the famous Cantor alloy as well as less known, but nevertheless important platinum group (PGM) alloys. By applying X-ray diffraction to samples loaded into diamond anvil cells we perform a comparative investigation of these HEA alloys in Ne and $H_2$ pressure-transmitting media. Surprisingly, even under stresses far exceeding conventional industrial processes both Cantor and PGM alloys show exceptional resistance to hydride formation, on par with widely used industrial grade Cu–Be alloys. Our observations inspire optimism for practical HEA applications in hydrogen-relevant industry and technology (e.g. coatings, etc), particularly those related to transport and storage.

# 1. INTRODUCTION

High-entropy materials, including high-entropy alloys (HEAs) and high-entropy oxides (HEOs), have gained significant attention as highly efficient applied materials [1]. Among other high-entropy solids, HEAs have simple crystal structures with long range crystal periodicity. At the same time, they have high structural and chemical disorder on the atomic level tightly correlated with exceptional properties. Originally developed as materials for structural applications, high-entropy materials were intensively studied as functional and energy-related materials with uses in catalysis and energy infrastructure (*e.g.* transport and storage). Considering the great variety of their properties, we find that high-entropy materials in general and HEAs in particular have a great potential to contribute to the ongoing developments in the vast field of hydrogen economy.

Indeed, binary and ternary metal hydrides were proposed as promising candidates for hydrogen storage applications in fuel cells. In comparison, refractory high-entropy alloy *bcc*–TiVZrNbHf can absorb much higher amounts of $H_2$ than its individual components and reach an H:*M* ratio of 2.5:1, with H and *M* corresponding to hydrogen and metal, respectively [2]. Such high $H_2$ content has never been observed in compositionally less complex interstitial hydrides based only on transition metals at ambient pressure and can be explained by the lattice micro strains in the alloy that make it favourable to absorb $H_2$ in both tetrahedral and octahedral interstitial sites. This observation published in 2016 stimulated finding of several Ti and Mg based HEAs with high $H_2$ uptake. Although, this is just a single example illustrating prospects for the energy storage, in reality, there are many other aspects of HEAs which have not been carefully explored (*e.g.* hydrogen uptake and potential for corrosion, hydrogen transport, direct applications including small and larger engine parts operating under elevated temperatures, *etc.*).

A review of the published literature signifies that only a few ternary alloys and HEAs have been characterized under compression in presence of $H_2$. Their phase transformations are considered to be much more complex in comparison with pure metals and binaries. For example, high local micro strains so typical for HEAs may allow alloy formation with significant $H_2$ uptake in their structure. Considering the perspectives of hydrogen storage, the review of the literature will highlight *bcc*-HEAs systems based on refractory metals with valence electron concentration (VEC) above 5 and with large lattice distortion [3]. The scarce theoretical models for Fe-based alloys suggest the following sequence for hydrogen diffusion coefficients: *bcc* >> *fcc* > *hcp* [4]. At the same time, we also see that behaviour of *hcp* and *fcc* alloys has not been characterized so far in great detail, but it is of great importance to understand hydrogen penetration between the closed-packed layers as it can also be significant. Hydrogen migration into octahedral and tetrahedral sites in closed-packed structures may have different energy and pathways resulting in a dramatical contrast of hydrogen uptake and hydrogen diffusivity. As a result, different mechanisms of hydrogen migration might be in play in *bcc*-, *fcc*- and *hcp*-structured alloys and they should be experimentally studied as function of composition and hydrogen pressure. A review of the literature indicates that the topic of hydrogen affinity as well as the topic of resistance to hydrogen with respect to HEA are greatly underexplored.

Among the great variety of HEAs, *fcc*–structured CoCrFeNiMn, also known as Cantor alloy, represents one of the most studied systems. It has been studied extensively at ambient pressure and under high pressure, with the latter studies typically conducted using inert pressure media.

From the chemistry point of view, the high-entropy alloy of *fcc*-structured CoCrFeNiMn as well as high entropy alloys based on platinum group (PGM) elements, e.g. *fcc*–Ir$_{0.231(2)}$Os$_{0.090(1)}$Pt$_{0.273(1)}$Rh$_{0.302(2)}$Ru$_{0.104(2)}$ and *hcp*–Ir$_{0.19(1)}$Os$_{0.22(1)}$Re$_{0.21(1)}$Rh$_{0.20(1)}$Ru$_{0.19(1)}$ described below, represent very contrasting examples to *bcc*–TiVZrNbHf and we selected these systems on a purpose.

On one hand, we would like to bring community's attention to the assortment of HEAs and the challenge of finding the most effective materials either for hydrogen related catalysis or hydrogen-resistant bulk material production and coatings (*e.g.* the quests of transport and storage). On the other hand, within our study we introduce a new quick testing methodology for material hydrogen resistance. Here, the cases of Cantor alloy and PGM HEAs allow us to illustrate different aspects of material behaviour under high pressure in hydrogen and importance of phase transitions for final analysis. Comparison of Cantor alloy and PGM HEAs up to 50 GPa seems appropriate. Indeed, under stress and even being surrounded by inert pressure transmitting medium (PTM), the *fcc*-phase of Cantor alloy can be reversibly converted to an *hcp*-structure. The transition occurs above 7 GPa and depends on a great number of factors, including features of the PTM and hydrostaticity of compression [5–7]. The PGM HEAs indicated above do not have any structural phase transitions and as we show below, their analysis is much simpler. Finally, our study would not be full without a comparative study of selected HEAs with industry relevant alloys including Alloy 190 and Toughmet 3 alloys produced by Materion®.

Technical development in hydrogen economy is extensive with its multifaceted aspects and applications on various levels (storage tanks, pipes, engines, etc.). We hope our study, apart from demonstrating a new efficient method for testing materials in hydrogen with a focus to hydrogen corrosion resistance in selected HEAs, will contribute to the overall progress and success in the field of renewable energy.

## 2. MATERIALS AND METHODS

Platinum group metals HEAs *fcc*–Ir$_{0.231(2)}$Os$_{0.090(1)}$Pt$_{0.273(1)}$Rh$_{0.302(2)}$Ru$_{0.104(2)}$ (*fcc*–PGM), *hcp*–Ir$_{0.19(1)}$Os$_{0.22(1)}$Re$_{0.21(1)}$Rh$_{0.20(1)}$Ru$_{0.19(1)}$ (*hcp*–PGM) were prepared as fine powders from single source precursors according to previously published protocols [8].

The CoCrFeMnNi HEA powder was produced at the Korea Institute of Materials Science (KIMS) by using vacuum induction gas atomizer (VIGA, HERMIGA 100/25, PSI, UK), and it is the same as used for additive manufacturing by laser power bed fusion, as described in detail in the publications [9–11]. A cast and homogenized ingot with the required equiatomic composition was first produced and then gas-atomized at 1580°C in an argon atmosphere. An inspection of the resulting CoCrFeMnNi material revealed a small grain polycrystals with dense microstructure, nearly equiatomic composition and a homogeneous distribution of the alloying elements [12,13].

Cantor alloy and PGM HEAs analytical composition was additionally probed and confirmed by scanning electron microscopy (SEM). SEM images were obtained on a XL30 ESEM (Environmental Scanning Electron Microscope) from FEI (Thermo Fisher Scientific, Waltham, Massachusetts, USA). The compositions were characterized by energy-dispersive X-ray analysis (EDAX, equipped with Si–(Li) detector) and averaged over 5–6 points. The averaged compositions were close to the nominal composition of single-source precursors and starting solutions. All the alloys reported in our paper are stable at ambient conditions.

For our comparative study, small metallic pieces of commercial alloys produced by Materion®, namely, Alloy 190 (*fcc*–A190) and Toughmet 3, (*fcc*–TM3) were cut out of bigger strips using a file. The reported composition of the Alloy 190 in weight percent is 1.8–2.0 wt.% Be, min. 0.2 wt.% Ni+Co, max. 0.6 wt.% Ni+Co+Fe with balance of Cu [14]. In turn, the ToughMet alloy is made of 15 wt.% of Ni, 8 wt.% of Sn and balanced with Cu [15]. Physical properties of these materials will depend on temper and, for both alloys, within the product nomenclature of Materion® we had materials of temper TM08 or XHMS. For additional details on physical properties of *fcc*–A190 and *fcc*–TM3 materials we refer to the their specifications [14,15].

Powder X-ray diffraction experiments using diamond anvil cells were conducted at two different synchrotron sources, namely, at ESRF-EBS in Grenoble and at PETRA III, DESY in Hamburg, and the respective extreme condition beamlines are ID15B and ECB/P02.2 [16]. In our study we employed diamond anvil cells equipped with 250 and 300 μm diamond culet size diameter. In **Table 1** we assemble the parameters of different high-pressure studies conducted at these facilities. We used ruby chips for pressure determination with sample loaded with He or Ne. Although the same ruby chips were used for pre-compression of diamond anvil cells loaded with $H_2$, we used Au for pressure determination during the compression for the latter loadings. All results presented in this manuscript were obtained from measurements at ambient temperature. Additional experimental details are shown in Supplementary.

Compressibility curves for *fcc*–CoCrFeMnNi Cantor alloy with non-reactive solid PTM (multi-anvil assembly) at moderate pressures below 15 GPa and room temperature were also collected using the large volume multi-anvil press Aster–15 installed at the P61B energy-dispersive beamline [17]. ~6 mg of *fcc*–CoCrFeMnNi Cantor alloy powder were pressed into a pellet (1.2 mm OD, ~0.75 mm height) and sealed inside a NaCl capsule (3 mm OD, ~3.4 mm height) along with 2 pellets of $NH_3BH_3$ (ammonia borane, 1.2 mm OD, ~0.75 mm height). We note that $NH_3BH_3$ is a very soft and non-reacting PTM if we consider ambient temperature compression [18], and that the sample capsule preparation was handled in an Ar-filled glove box. The sample was compressed using a 14/7 multi-anvil assembly, which is described in detail elsewhere [19]. The octahedron was further positioned between eight 32 mm WC anvils (Fujilloy TF08, 7 mm TEL) equipped with pyrophyllite gaskets. Pressures were estimated using the NaCl equation of state [20]. Here the powder diffraction data was collected up to the maximum pressure was 15 GPa.

**Table 1** High-pressure X-ray diffraction on different alloys.

| Facility | Beamline | Alloy | PTM | Detector | Wavelength, Å | Beam size at sample, FWHM, μm² |
|---|---|---|---|---|---|---|
| ESRF-EBS | ID15B | *fcc*–PGM | He | EIGER2 | 0.4103 | 10·10 |
| ESRF-EBS | ID15B | *hcp*–PGM | Ne | MAR555 | 0.4110 | 10·10 |
| ESRF-EBS | ID15B | *fcc*–Cantor | He & $H_2$ | EIGER2 | 0.4103 | 10·10 |
| DESY | P61B | *fcc*–Cantor | LVP | Ge-SSD by Mirion (Canberra) for ED-XRD positioned at 4.962° and 3.038° 2θ | energy dispersive (30–160 keV) | 50·300 |
| DESY | P02.2 | *fcc*–PGM | $H_2$ | PE | 0.2907 | 2·2 |
| DESY | P02.2 | *hcp*–PGM | $H_2$ | PE | 0.2907 | 2·2 |
| DESY | P02.2 | *fcc*–A190[1] | Ne & $H_2$ | PE | 0.2909 | 2·2 |
| DESY | P02.2 | *fcc*–TM3[1] | Ne & $H_2$ | PE | 0.2909 | 2·2 |

FWMH – full width at half-maximum, H·V; PTM – pressure transmitting medium; PE – detector Perkin Elmer 1621 XRD; EIGER2 - EIGER2 X 9M CdTe; [1] – both samples were measured at the same time either in a single Ne or single $H_2$ loading; LVP – Large volume press assembly including NaCl+ $NH_3BH_3$.

The collected 2D diffraction patterns were integrated using DIOPTAS software [21]. The unit cell parameters, the background and the line-profile parameters were simultaneously refined using TOPAS, JANA2006 and GSAS-II software [22–24]. The data and the corresponding equations of state were analysed using EOSFIT7-GUI [25,26]. Energy-dispersive X-ray diffraction patterns were initially visualized and evaluated using PDindexer software tool [27].

## 3. RESULTS AND DISCUSSION
### 3.1. AFFINITY OF *fcc*–A190 *fcc*–TM3 to HYDROGEN UNDER COMPRESSION

Before starting the presentation of high entropy alloys data, we show our results on conventional industrial alloys, namely, *fcc*–A190 and *fcc*–TM3 as compressed in Ne and $H_2$.

We will first have a look at data obtained with Ne PTM shown in **Figure 1**. Our data suggest absence of any phase transitions *fcc*-A190 and *fcc*–TM3 under compression at ambient temperature. Since both alloys are based on copper, thus, it is not surprizing that their corresponding equations of state (EOS) closely follow the *V–P* curve of pure *fcc*–Cu. The contrast of absolute volume values is likely dependent on alloy doping elements. In **Figure 1** we show the data fit to the 3$^{rd}$ order Birch-Murnaghan EOS [28], where the parameters $V_0$, $K_0$ and *K'* correspond to ambient pressure volume, ambient pressure bulk modulus and its pressure derivative, respectively.

The picture won't change if Vinet EOS is employed. In latter case the parameters are $V_0$=11.56(2) Å$^3$, $K_0$=138(9), K'=4.9(8) for *fcc*-A190 and $V_0$=11.97(1) Å$^3$, $K_0$=147(6), K'=4.9(6) for *fcc*–TM3, respectfully. Considering the *fcc*–Cu Vinet EOS data published in [29], we can refer to $V_0$=11.81 Å$^3$, $K_0$=135(1), K'=4.91(5) for the purpose of comparison. Although, the compressibility of solids will also depend on temper of the alloys (sintered grains), in general we see a good agreement with pure *fcc*–Cu considering that alloys *fcc*–A190 and *fcc*–TM3 have major abundance of Cu.

After we determined EOS describing the behaviour of *fcc*–A190 and *fcc*–TM3 EOS under quasi-hydrostatic conditions of Ne, we can compare this data with behaviour of the same alloys in $H_2$ PTM. This comparison lays a foundation for a simple method testing industry-relevant materials for their affinity to hydrogen, and we introduce the procedure here.

Quasi-hydrostatic compression of materials sets a reference line, indeed, for a given pressure value the unit cell volume may not be smaller than the equilibrated reference, if we consider cthermodynamic equilibrium. Formation of hydrides often occurs at ambient temperature under sufficient stress and may result either in (A1) penetration of original crystal structure and formation of an isostructural hydride or in (A2) formation of hydrides with the lower pressure structure undergoing a transition under compression. For simple *fcc*–A190 and *fcc*–TM3 we did not expect any phase transitions below 35 GPa (**Figure 1**), thus, they represent a great case study for (A1).

Our data on compression and decompression cycles of *fcc*–A190 and *fcc*–TM3 are indeed indicative (**Figure 2**). We can easily attribute the regions of material resistance to hydrogen and the regions of hydride formation. We observed isostructural hydride formation resulting in an increase of the unit cell volumes and the volumes measured per atom. Unambiguously, *fcc*–A190 is superior to *fcc*–TM3 as indicated by the higher pressures of hydride formation and the excess volume per atom within the hydride stability region. It is also worth to note, that in systems like *fcc*–A190 and *fcc*–TM3 amount of hydrogen is variant with pressure. Our data collected on decompression indicates higher volume of *fcc*–TM3 for the indicated pressure in comparison to the compression run (*e.g.* point ~33 GPa). The *fcc* and *hcp* lattices of conventional alloys and the PGM HEAs demonstrated below have similar interstitial voids for hydrogen to occupy. These voids have tetrahedral and octahedral local environments and are occupied depending on atomic size of constituent metals. For example, transition elements with larger and smaller atomic sizes likely capture hydrogen in tetrahedral sites and octahedral sites, respectively [30].

Our case study involving a combination of diamond anvil cell with X-ray diffraction with the application to the industry relevant alloys of *fcc*–A190 and *fcc*–TM3 suggests that the proposed here method of testing materials and comparing their compression in quasihydrostatic PTM and in corrosive PTM, such as $H_2$, has a great potential. It has also several advantages for sample screening: 1) it requires negligible amount of test material; 2) several samples can be tested at the same time; 3) sample chamber is small and the experiment is less dangerous in comparison to large autoclaves; 4) experiment is fast. Indeed, a single slow compression with pressure controlled by a gas driven membrane with several samples per sample chamber may take 3–6 hours for data acquisition for a single PTM with controllable strain rate. Using dynamic piezoelectric actuated diamond anvil cells, screening times can be significantly shortened, resulting in a full data acquisition within minutes if not seconds [31].

### 3.2. COMPRESSION OF HIGH ENTROPY ALLOYS

Using the *fcc*–A190 and *fcc*–TM3 as an example, we described our observations and typical behaviour of alloys compression in quasi-hydrostatic and $H_2$ PTM in the case (A1) when hydrogen penetrates crystal lattice and induces an isostructural formation of hydrides.

Our *fcc*–PGM and *hcp*–PGM alloys were behaving the similar way as *fcc*–A190 and *fcc*–TM3 (**Figure 3**). Surprisingly, the studied PGM based HEA alloys are also resistant to hydrogen upon compression. Comparing both systems with one another we suggest that *fcc*–PGM is less resistant than *hcp*–PGM with the lower starting pressure of hydride formation and the overall larger capacity.

We can approach the challenge of hydrogen content estimation in *fcc*- and *hcp*-structured PGM HEAs similarly to the approach published by Somenkov *et al*. [30]. Following their idea and experimental observations, we consider interstitial hydrogen atoms as incompressible and interacting with free electrons of the metal (*M*). Assessing the formation of $MH_x$ hydride as a high entropy hydride, we can roughly estimate the hydrogen concentration as $x = \Delta V / \delta v$, where

*ΔV* is excess volume of *MH$_x$* over pure *M* for its single atom, and *δv* is a parameter depending on various factors, including the number of valence electrons attributed to *M*. For the full description we refer to original discussion [30]. We calculate *δv* of HEAs as a linear interpolation of individual constituents and for *fcc*–PGM and *hcp*–PGM, we calculated the *δv* values as equal to 2.368 Å$^3$ and 2.112 Å$^3$, respectively.

The resulting data is shown in **Figure 4.** The compression trend indicates that pressure as high as 50 GPa does not saturate the *hcp*–PGM with hydrogen. Indeed, the compression slope of *hcp*–PGM does not even start to decrease with compression.

Based on our data we can arrange the materials with respect to their affinity to hydrogen in the following way, with the least resistant mentioned first: 1) *fcc*–TM3; 2) *fcc*–PGM, and 3) *hcp*–PGM and *fcc*–A190. Here, we would like to avoid an impression that these materials can be used for the same applications. Our study only indicates that like in the case of industrial materials, affinity of HEAs to hydrogen is controlled by their crystal chemistry (controlled by doping) and can be efficiently tested using the combination of the diamond anvil cell and X-ray diffraction techniques. It also indicates compression induced formation of previously unknown high entropy hydrides with PGM HEA systems as precursors.

Last, but not least we present the data on compression of Cantor alloy in hydrogen PTM. In contrast to the previously mentioned conventional and high-entropy alloys, Cantor alloy belongs to the case (A2) and undergoes a phase transition at relatively low pressures. The transformation sluggish, and it is well described in literature [6,32,33].

Our results of Cantor alloy compression are in good agreement with literature (**Figure 4**). The small deviations observable at pressures above ~25 GPa for *hcp* and *fcc* phases for material compressed in diamond anvil cell could be explained by contributions from non-hydrostatic effects (*e.g.* see Glazyrin *et al.* [34] for more details). These deviations indicate that for sintered grain materials the effects of non-hydrostaticity may develop even if one uses He as the most hydrostatic pressure medium, especially if a sample has sintered grains and if it undergoes a martensitic phase transition (stronger influence of inter-grain stress and intrinsic micro-strains). Still, although such effects exist, and we should be aware of these details, the most important observations and conclusions remain unchanged.

Here, we attract the attention to the following important observation. If we compare data obtained under ambient temperature compression in large volume press PTM, silicone oil, He and H$_2$ pressure media we conclude that there is only minimal absorption of hydrogen by the Cantor alloy in contrast to *fcc*–TM3 presented above and, moreover, to the 316 Steel also widely used for hydrogen-technology as our results show in Supplementary.

Our data shows that the low pressure *fcc*-phase of Cantor alloy does not absorb hydrogen up to the pressures of *hcp*-phase transition which according to Tracy et al. should start at 14 GPa. This result indicates that for certain applications Cantor alloy could be even better than *fcc*–A190 containing Be and, thus, more environmentally friendly in terms of recycling, *etc*.

We deem these results open a new venue to hydrogen economy research and development, particularly within the sub-fields of hydrogen production and storage, transportation, and infrastructure. They are simultaneously surprising and promising and hold potential for various industrial HEA applications.

## CONCLUSIONS

The research and development within the initiative of green energy and sustainability is ongoing and requires intensive exploration along multiple fronts: *e.g.* physics, chemistry, material science *etc*. The increasing trend in hydrogen economy investments is evident improving the aspects of hydrogen production as well as the realms of hydrogen transportation and storage.

In our work we attract the community attention to the new fast characterization method requiring small amount of sample to test material affinity of reaction with hydrogen. We introduce the method and apply it to the industrial and novel materials, such as high entropy alloys, and give a qualitative report for individual material affinity to hydrogen and discuss importance of composition in correlation with absorbed hydrogen content.

Considering the PGM HEAs we indicate an intriguing fact of their low hydrogen affinity despite large atomic radii of constituent atoms. These materials could be important for catalytical applications.

Finally, but not least, we investigate the Cantor alloy – the HEA alloy with probably the largest number of potential applications and find that it also is very resilient in comparison to 316 Steel as well as to Cu–Be alloy. Our results indicate that Cantor HEA is the same level and likely more efficient in comparison to the many applied materials widespread within the fields of hydrogen economy. Considering that the price of material production decreases and the scope of applications broadens with time and with technological progress, we hope to see the first implementations of Cantor alloy in various fields of hydrogen economy very soon.

Cantor alloy represents a single example within the broad space of HEA compositions, the space encompassing a large variety of physical and chemical properties. The number of potential applications for Cantor alloy is very exciting,, but our work fills only one piece of a bigger puzzle. Here, we explore HEA resilience from a new perspective, and, among other results, show formation of new high entropy hydrides in previously unexplored systems, but it is just a drop in the ocean. We hope that our work will attract the community attention to the vast and still unexplored space, as the more complete picture of HEAs will drive the technological process further and farther opening new avenues for innovation.

## ACKNOWLEGEMENTS

We acknowledge DESY (Hamburg, Germany), a member of the Helmholtz Association HGF, for the provision of experimental facilities. Parts of this research were carried out at beamlines P02.2 and P61B. We also thank the ID15B beamline at the European Synchrotron Radiation Facility, Grenoble, France, for providing us with the measurement time and technical support. A partial support from German Research Foundation (DFG) via research grant DI 1419/24-1 is acknowledged. We appreciate and acknowledge the help and advice of Klaus Ohta, Magdalini Dalla and Andreas Frehn from Materion®s providing test *fcc*-A190 and *fcc*-TM3 materials.ssssssss

## FIGURES

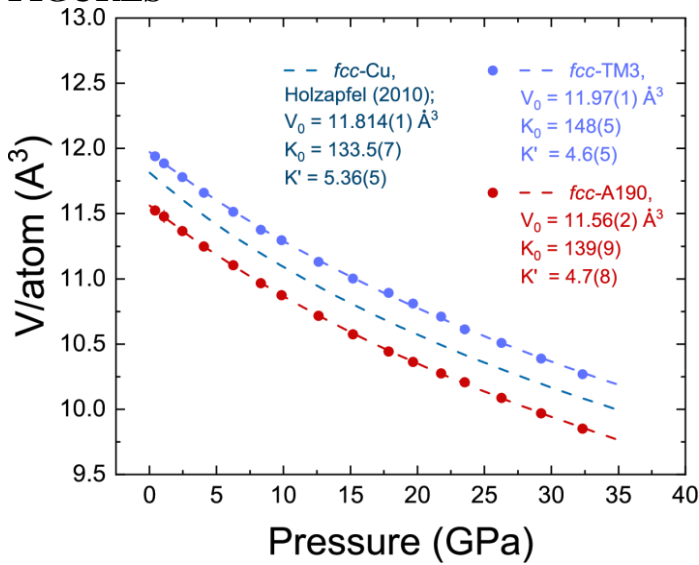

**Figure 1** Equations of state (EOS) for pure *fcc*–Cu, *fcc*–A190 and *fcc*–TM3 alloys shown as unit cell volume normalized by number of constituent atomic sites. The experimental data is shown using the symbols and the fits to 3$^{rd}$ order Birch-Murnaghan EOS (BM3) are shown as dashed lines. If not visible, the error bars are the size of the symbol or below. For *fcc*–Cu we use reference [35]. It is clear that compression of the alloys is governed by Cu constituent, although the starting volumes differ due to Be (smaller atom in *fcc*–A190) and Sn (larger atom in *fcc*–TM3) doping.

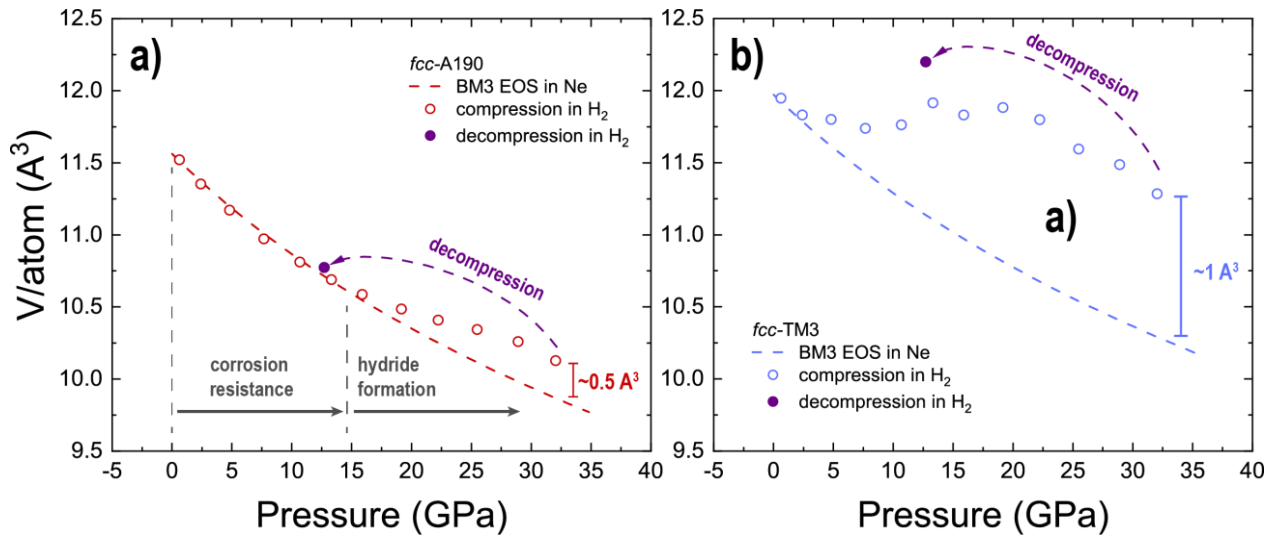

**Figure 2** Comparison of unit cell volume compression normalized to the number of atoms per unit cell for a) *fcc*–A190 and b) *fcc*–TM3. The red and blue dashed lines represent the corresponding BM3 EOS quasi-hydrostatic Ne PTM as shown in **Figure 1**. The open circles correspond to data measured upon compression in $H_2$. The purple solid symbol represents decompression data. In a) we split the compression curve in several regions. Within the corrosion resistance field, the volume per atom of *fcc*-A190 compressed in $H_2$ follows the compression line measured in Ne. However, starting from 15 GPa we see an increase of *fcc*–A190 attributed to hydride formation. It is clear from b) that *fcc*–TM3 has higher affinity to $H_2$ than *fcc*–A190. The data on decompression for both alloys suggests that $H_2$ content increases with compression and indicates suppressed kinetics, *e.g.* for hydrogen it is easier to diffuse into the lattice upon compression and harder to escape from crystal lattice of alloy if pressure is subsequently reduced.

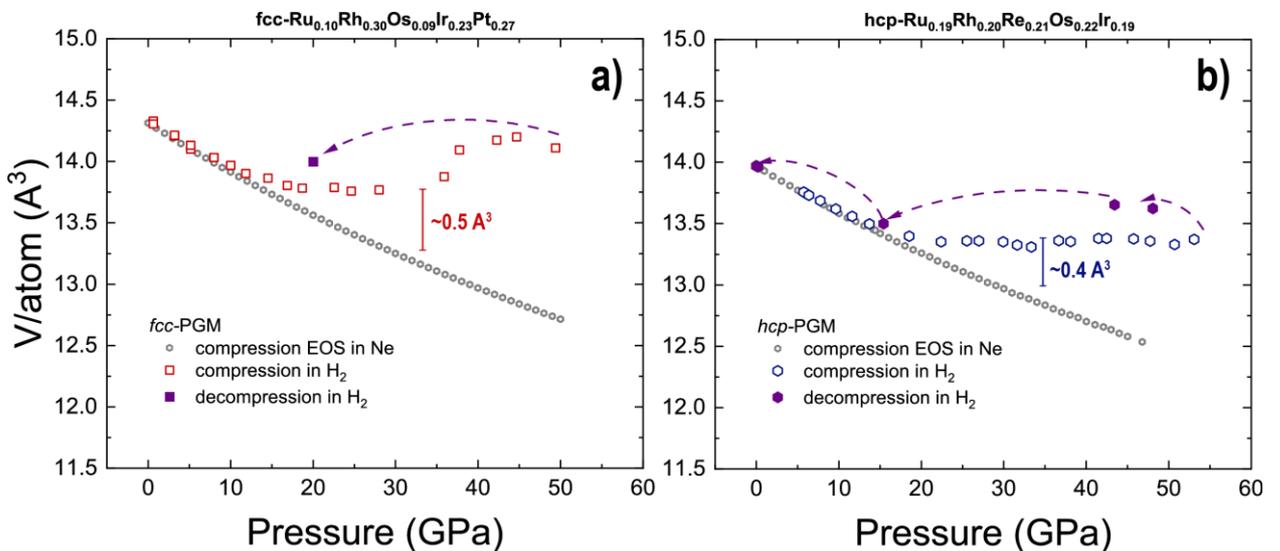

**Figure 3** Compression curves of a) *fcc*–PGM and b) *hcp*–PGM. Grey open hexagons represent the corresponding EOS'es measured with Ne pressure medium. Despite being formed from atoms with large atomic radii, both alloys exhibit considerable resistance to $H_2$ with *hcp*–PGM being the most resistant from the two. Considering the pressure point of 30 GPa, the excess volume of *hcp*–PGM attributed to $H_2$ capture is similar to *fcc*–A190 and has a value of ~0.4 Å$^3$/atom. Decompression trend suggests suppressed kinetics and trapping of hydrogen on decompression, but full release of $H_2$ at ambient conditions. The red and blue bars indicate approximate volume difference with respect to compression of the same material in Ne for pressures in the vicinity of 30 GPa. The parameters for the HEAs 3$^{rd}$ order Birch-Murnaghan EOS are: $V_0/Z = 14.18(2)$ Å$^3$/atom, $K_0 = 297(3)$ GPa, $K' = 5.8(1)$ GPa for *fcc*–PGM

(compression in He) and $V_0/Z = 13.979(2)$ Å$^3$/atom, $K_0 = 317(2)$ GPa, $K' = 4.9(1)$ GPa for *hcp*–PGM (compression in Ne).

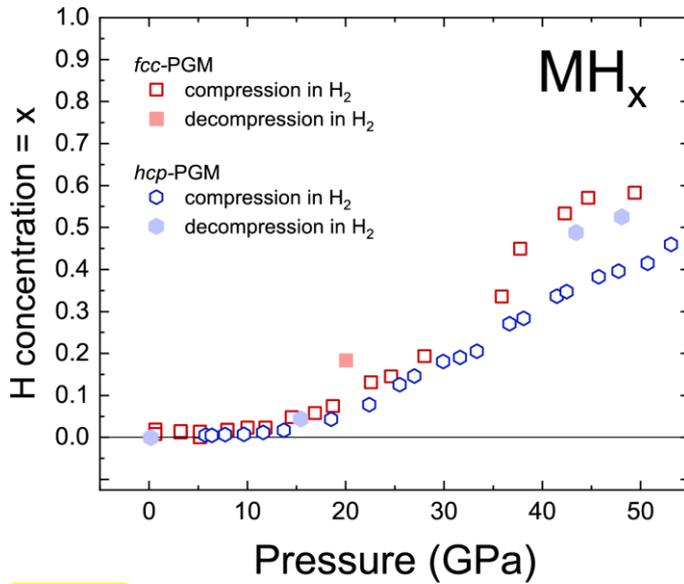

**Figure 4** Estimation of hydrogen content of *fcc*– and *hcp*–PGM HEAs according to methodology of Ref. [30]. We assume formation of $M$H$_x$ where $M$ is metal atom, H is hydrogen, and $x$ stands for its concentration in hydride. Potentially, in contrast to *hcp*–PGM, *fcc*–PGM captures small amount of hydrogen already at pressures below 10 GPa indicating importance of composition for material performance.

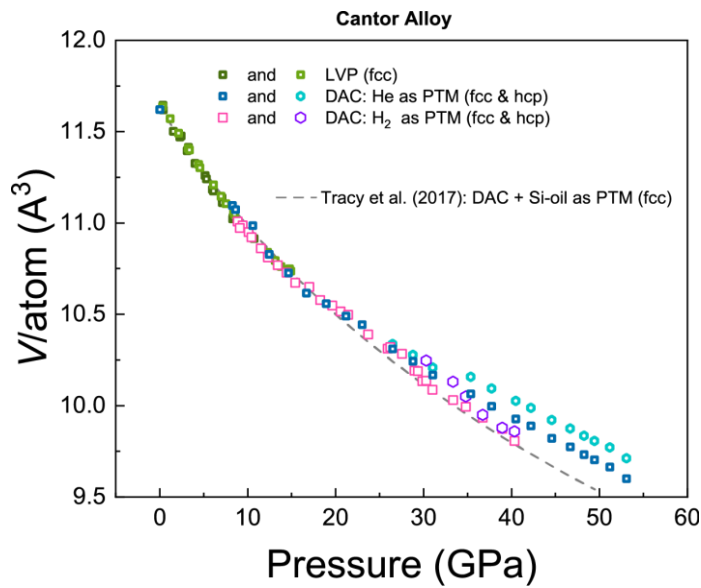

**Figure 5** Cantor alloy compression data. We show the data from the large volume press together with the data collected from diamond anvil cell experiments and indicate the corresponding phases. For comparison we show the equation of state reported by Tracy *et al*. [32] by means of the grey dashed line. Our data is in a good agreement with previously reported results. Note how close the individual phases of Cantor alloy compressed under H$_2$ PTM follow the equation of state of Tracy *et al*. up to the highest measured pressures.

# SUPPLEMENTARY

High entropy alloys and their affinity to hydrogen: from Cantor to platinum group elements alloys.

K. Glazyrin,[a] K. Spektor,[a] M. Bykov,[b] W. Dong,[a] J.-H. Yu,[c] S. Yang,[c] J.-S. Lee,[d] S. Divinski,[e] M. Hanfland,[f] K. V. Yusenko[g]

[a] *Photon Sciences, Deutsches Elektronen-Synchrotron, Notkestr. 85, 22607 Hamburg, Germany,* [b] *Institute of Inorganic Chemistry, University of Cologne, Cologne, Germany,* [c] *Powder Materials Division, Korea Institute of Materials Science, 51508 Changwon, South Korea,* [d] *Department of Materials Science and Chemical Engineering, Hanyang University, 15588 Ansan, South Korea,* [e] *Institute of Materials Physics, University of Münster, D-48149 Münster, Germany,* [f] *ESRF – The European Synchrotron, 71 Av. des Martyrs, 38000, Grenoble, France,* [g] *Bundesanstalt für Materialforschung und –prüfung (BAM), D-12489 Berlin, Germany*

## 1. Compression of industrial alloys *fcc*–A190, *fcc*–TM3 and 316 Steel

Here and below we show some additional information with respect to the industrial alloys. In addition to the study on *fcc*–A190 and *fcc*–TM3 (**Figure S1**) we also conducted a study on conventional 316 Steel purchased from GoodFellow® [1]. From the point of X-ray diffraction, this steel material can be described as two-phase composition: the martensitic (*bcc*) and the austenitic (*fcc*) phases (**Figure S2**).

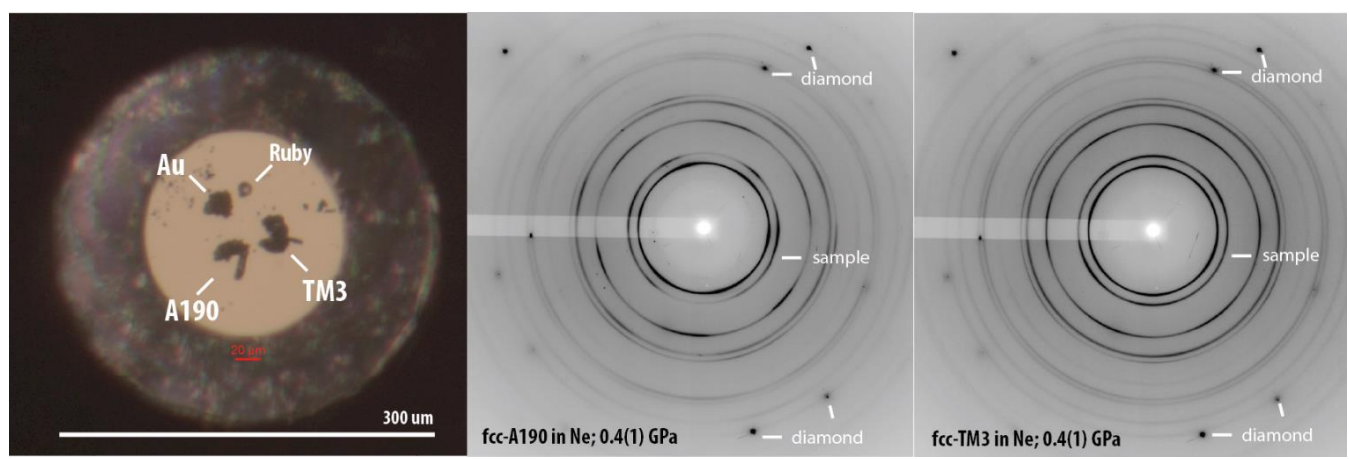

**Figure S1** Images of (left) microphotograph of a diamond anvil cell sample chamber prior to hydrogen loading. It shows *fcc*–A190 (Alloy 190), *fcc*–TM3 (ToughMet 3) as well as chips of ruby and gold. Ruby was used for pre-compression after gas-loading and gold was used for pressure determination during compression in $H_2$. The images to the (right) and in the (middle) show 2D diffraction images of the alloys loaded with Ne. The same images confirm abundant presence of *fcc*-structured phases with preferred orientation of the grains due to manufacturing process.

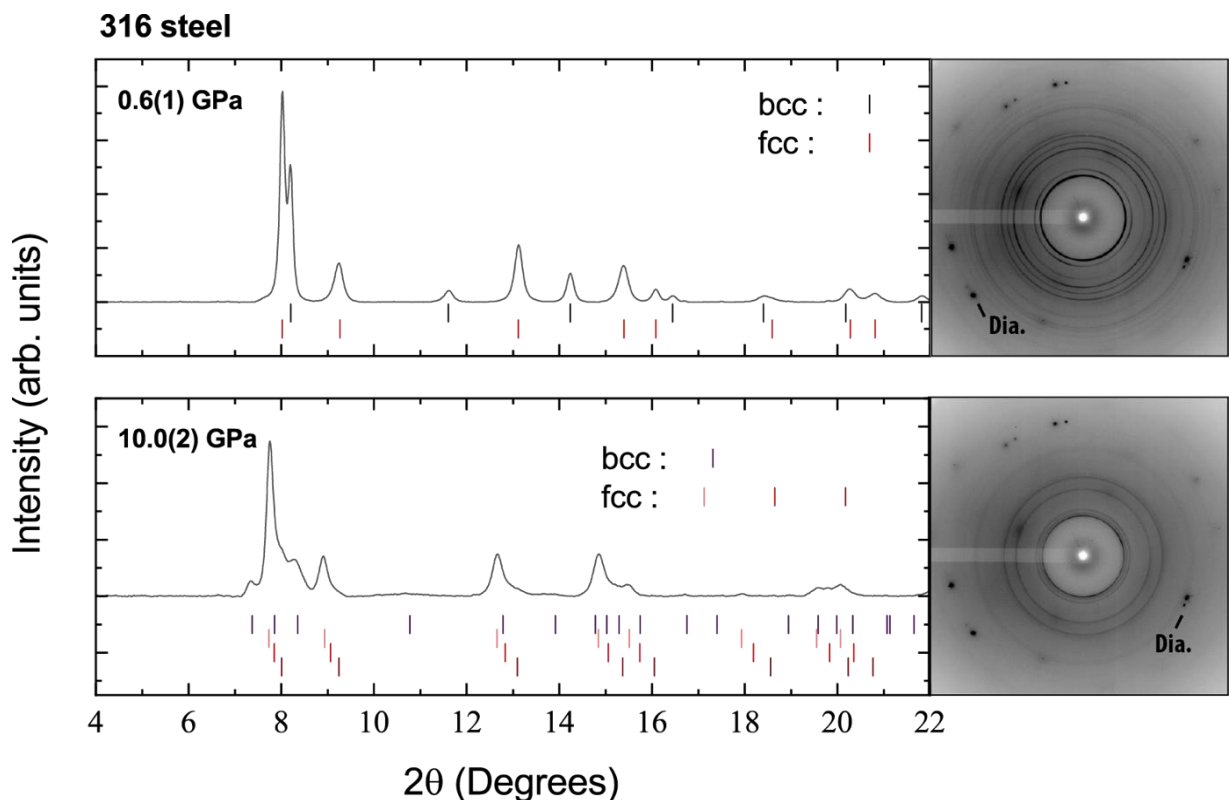

**Figure S2** Selected diffractograms for 316 steel compression in H$_2$ pressure medium. Under compression we see formation of and *hcp, fcc* hydrides with different hydrogen content, pressure dependent. Considering the *fcc* phase, we see slow capture of hydrogen, with intermediate hydrogen content evident through the tails of *fcc* phase peak for 10.0(2) GPa diffractogram. We indicate peaks from diamond anvils at the corresponding 2D image.

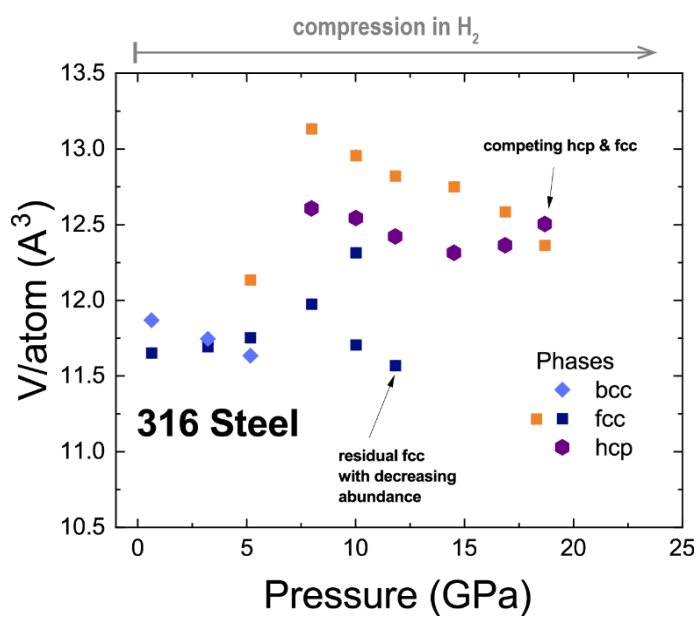

**Figure S3** Atomic **v**olume for phases measured upon compression of steel ALSI 316 [1] in hydrogen at room temperature. The *fcc* and *bcc* phases present at ambient are converted to hydrides at relatively low pressure. At pressures above 5 GPa we could not observe *bcc* as it transforms to the *hcp* phase. The positive trend of *fcc* phase volume as a function of pressure range below 10 GPa indicates that this phase starts to pick up hydrogen early.

Considering the data collected from 316 steel we see that hydrogen is captured by the material already at low pressures in contrast to *fcc*–A190. Considering the chemical composition, the steels 304, 316 and 316L steels are not that different. This difference is very important for their mechanical properties after tempering acid resistance (higher for 316 and 316L), but we consider that, although some details will be different, in general they will behave very similar under compression with hydrogen.

## 2. Compression of *fcc*–PGM and *hcp*–PGM

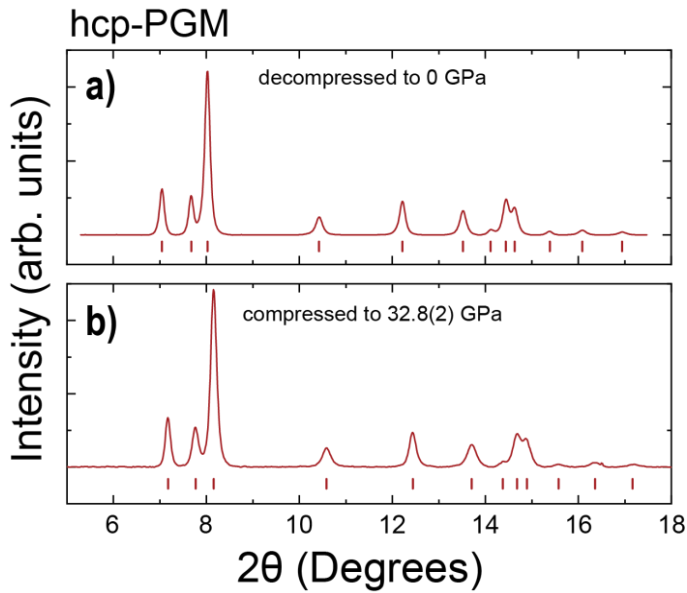

**Figure S4** X-ray diffraction patterns collected on *hcp*–PGM (composition $Ir_{0.19(1)}Os_{0.22(1)}Re_{0.21(1)}Rh_{0.20(1)}Ru_{0.19(1)}$). In a) we show data collected after decompression in $H_2$ pressure transmitting medium (PTM) to ambient pressure while in b) we show diffraction patterns after compression in $H_2$ PTM with finite amount of hydrogen captured within the crystal lattice of the compound.

## 3. Compression of Cantor Alloy

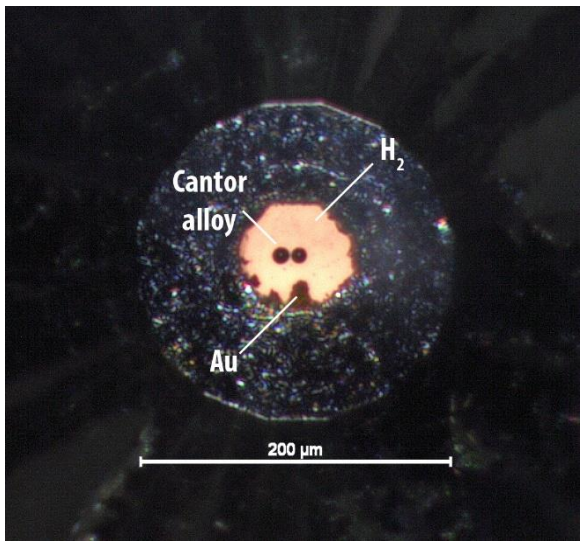

**Figure S5** Microphotograph showing small spheres of Cantor alloy loading with hydrogen as pressure-transmitting medium. We used powder Au with sub-micron grain as pressure standard.